\documentclass[conference,letterpaper]{IEEEtran}
\IEEEoverridecommandlockouts

\usepackage{amsmath,amssymb,amsfonts}
\usepackage{algorithmic}
\usepackage{graphicx}
\usepackage{textcomp}
\usepackage[OT1]{fontenc}
\usepackage{pifont}
\usepackage{multirow}
\usepackage{threeparttable}
\usepackage{placeins}
\usepackage{booktabs}
\usepackage{amsmath}

\usepackage[dvipsnames,svgnames]{xcolor, colortbl}

\definecolor{LightCyan}{rgb}{0.88,1,1}

\def\BibTeX{{\rm B\kern-.05em{\sc i\kern-.025em b}\kern-.08em T\kern-.1667em\lower.7ex\hbox{E}\kern-.125emX}}
\bstctlcite{IEEEexample:BSTcontrol}

\renewcommand\normalsize{\fontsize{11pt}{13pt}\selectfont}

\usepackage[bookmarks=false,hidelinks]{hyperref}

\begin{document}

\title{Efficient and Scalable Self-Healing Databases Using
 Meta-Learning and Dependency-Driven Recovery}

\author{\IEEEauthorblockN{Joydeep Chandra}
\IEEEauthorblockA{\textit{BNRIST, Dept. of CST} \\
\textit{Tsinghua University}\\
Beijing, China}
\and
\IEEEauthorblockN{Prabal Manhas}
\IEEEauthorblockA{\textit{Department of Informatik} \\
\textit{Technische Universität Kaiserslautern}\\
Kaiserslautern, Germany}
}

\maketitle

\begin{abstract}
Modern database management systems (DBMS) face significant challenges in maintaining performance and availability under dynamic workloads. This paper proposes a novel self-healing framework that integrates Model-Agnostic Meta-Learning (MAML) for few-shot anomaly detection, Graph Neural Networks (GNNs) for dependency-driven cascading failure prediction, and multi-objective Reinforcement Learning (RL) for autonomous recovery. Unlike existing database tuning systems that focus primarily on offline configuration optimization, our framework enables real-time, end-to-end self-healing by rapidly adapting to unseen workload patterns with minimal labeled data. We introduce dynamic GNN-based dependency modeling that captures workload-dependent relationships between database components, enabling proactive cascade prevention. A scalarized multi-objective RL formulation balances latency, resource utilization, and cost during recovery, while SHAP-based explainability ensures operational transparency. Evaluations on Google Cluster Data and TPC benchmarks demonstrate 90.5\% anomaly detection F1-score with 5-shot adaptation, 90.1\% cascade prediction accuracy, and 85.1\% latency reduction in recovery actions, outperforming strong baselines including Isolation Forest, LSTM autoencoders, static GCN, and standard RL methods.
\end{abstract}

\begin{IEEEkeywords}
Self-Healing Databases, Meta-Learning, Model-Agnostic Meta-Learning (MAML), Anomaly Detection, Graph Neural Networks (GNNs), Reinforcement Learning (RL), Multi-Objective Optimization, Cascading Failure Prediction, Explainable AI (XAI), Database Management Systems (DBMS).
\end{IEEEkeywords}

\section{Introduction}
\label{sec:Introduction}

Modern database management systems (DBMS) face unprecedented challenges in maintaining performance and availability under dynamic workloads, particularly in cloud-based and distributed environments. Real-world production systems experience anomalies ranging from resource contention and cascading failures to query performance degradation, often requiring manual intervention and lengthy retraining of detection models \cite{Pavlo2017SelfDriving, Cheng2023AIOps}. The demand for ``self-driving'' database systems that can autonomously detect, diagnose, and recover from failures with minimal human intervention has become critical in modern data centers \cite{Pavlo2021SelfDriving, Zhou2021DBMind}.

Prior research has explored various machine learning approaches to address these challenges. Bayesian Optimization (BO) based systems such as OtterTune \cite{VanAken2017OtterTune} leverage Gaussian Processes and workload mapping to recommend knob configurations, while iTuned \cite{Duan2009iTuned} pioneered automated parameter tuning using iterative experimentation. More recently, Deep Reinforcement Learning (DRL) has been applied for database configuration tuning: CDBTune \cite{Zhang2019CDBTune} uses Deep Deterministic Policy Gradient (DDPG) for end-to-end cloud database tuning, QTune \cite{Li2019QTune} extends this with query-aware Double-State DDPG for fine-grained tuning, and UDO \cite{Wang2021UDO} proposes universal database optimization using RL with delayed feedback for heavy parameters. ResTune \cite{Zhang2021ResTune} further introduces meta-learning to accelerate constrained Bayesian Optimization for resource-oriented tuning.

However, these existing approaches exhibit fundamental limitations when applied to \textit{self-healing} scenarios:
\begin{itemize}
    \item \textbf{Offline Optimization Bias}: OtterTune, CDBTune, and UDO require lengthy offline tuning sessions (15--45 minutes or hours) with iterative workload replay, making them unsuitable for real-time anomaly recovery \cite{VanAken2021Inquiry}.
    \item \textbf{Poor Workload Adaptation}: Traditional DRL methods such as CDBTune suffer from lengthy convergence cycles and poor generalization to unseen workloads; they require complete retraining when workload characteristics shift \cite{Zhang2021CDBTunePlus}.
    \item \textbf{No End-to-End Recovery}: Learned query optimizers such as Neo \cite{Marcus2019Neo} and Bao \cite{Marcus2021Bao} improve plan selection but do not address anomaly detection, failure prediction, or autonomous recovery actions.
    \item \textbf{Lack of Dependency Awareness}: Existing systems treat database components in isolation, missing complex interdependencies that drive cascading failures across queries, tables, and indexes \cite{Chen2009CascadeFailures}.
\end{itemize}

Meta-learning offers a paradigm shift in addressing the adaptability gap. Model-Agnostic Meta-Learning (MAML) \cite{Finn2017MAML} enables models to rapidly adapt to new tasks with minimal training data through few-shot learning principles. This capability is particularly valuable for database systems where labeled anomaly examples are scarce and workload patterns evolve continuously \cite{Hospedales2021MetaLearning}. Complementary approaches include REPTILE \cite{Nichol2018REPTILE} for simpler first-order meta-learning and Prototypical Networks \cite{Snell2017PrototypicalNetworks} for metric-based few-shot adaptation.

A key challenge in self-healing databases is modeling complex interdependencies between system components. Graph Neural Networks (GNNs) \cite{Kipf2017GCN, Hamilton2017GraphSAGE} have emerged as powerful tools for representing relational structures, with applications in learned query optimization \cite{Marcus2019Neo, Marcus2021Bao} and system dependency modeling \cite{Wu2021GNN}. For cascading failure prediction, GNNs can explicitly model how anomalies propagate from stressed components (e.g., lock contention on a hot table) to dependent queries and downstream services \cite{Chen2009CascadeFailures}.

This paper proposes an integrated self-healing framework that combines:
\begin{enumerate}
    \item \textbf{MAML-based anomaly detection} for rapid adaptation to new workload anomalies with 5--20 labeled examples;
    \item \textbf{Dynamic GNN dependency modeling} for real-time cascading failure prediction;
    \item \textbf{Multi-objective RL recovery} that dynamically balances latency, resource utilization, and cost;
    \item \textbf{SHAP-based explainability} \cite{Lundberg2017SHAP} for transparent decision-making; and
    \item \textbf{Synthetic task augmentation} via generative models \cite{Goodfellow2014GAN} to improve data efficiency in sparse regimes.
\end{enumerate}

Our work makes the following contributions:
\begin{itemize}
    \item We present the first integrated framework combining MAML, dynamic GNNs, and multi-objective RL specifically for end-to-end database self-healing, distinguishing our approach from existing offline configuration tuners.
    \item We introduce a dynamic GNN architecture with workload-aware edge weights that improves cascade prediction accuracy by 6--8 percentage points over static GCN baselines.
    \item We demonstrate that MAML enables 100$\times$ reduction in adaptation steps compared to full retraining, achieving 87.9\% F1-score with only 5 labeled examples on unseen workloads.
    \item We provide a comprehensive empirical evaluation against seven verified baselines on real-world datasets, including an ablation study and hyperparameter sensitivity analysis.
\end{itemize}

The remainder of this paper is organized as follows. Section~\ref{sec:Background} reviews related work and defines state-of-the-art baselines. Section~\ref{sec:Methodology} details the proposed framework. Section~\ref{sec:Results_and_Analysis} presents experimental results. Section~\ref{sec:Conclusion} concludes with limitations and future directions.

\section{Related Work and Baseline Methods}
\label{sec:Background}

We organize related work into four categories: (1) self-driving database architectures and configuration tuning, (2) anomaly detection for database systems, (3) graph-based dependency modeling, and (4) meta-learning and few-shot adaptation. For each category, we define verified baselines and their limitations, establishing the foundation for our proposed framework.

\subsection{Self-Driving Databases and Configuration Tuning}

The vision of autonomous DBMS administration has been pursued for over a decade. Chaudhuri and Narasayya \cite{Chaudhuri2007SelfTuning} surveyed early self-tuning techniques including automatic index selection and materialized view maintenance. More recently, Pavlo et al.\ \cite{Pavlo2017SelfDriving, Pavlo2021SelfDriving} outlined architectural principles for self-driving DBMS encompassing workload forecasting, automatic physical design, and runtime adaptation.

\textbf{Baseline BO-based Tuning:} OtterTune \cite{VanAken2017OtterTune} is the most widely cited BO-based database tuning system. It uses LASSO for knob importance ranking, workload fingerprints for similarity matching, and Gaussian Process regression for configuration recommendation. iTuned \cite{Duan2009iTuned} and BestConfig \cite{Zhu2017BestConfig} represent earlier heuristic and direct-search approaches. Limitations: these methods require 15--45 minutes of iterative workload replay per tuning session, assume static workloads, and provide no real-time anomaly recovery \cite{VanAken2021Inquiry}.

\textbf{Baseline RL-based Tuning:} CDBTune \cite{Zhang2019CDBTune} applies DDPG for end-to-end cloud database knob tuning with a trial-and-error strategy. QTune \cite{Li2019QTune} extends CDBTune with query-aware DS-DDPG, supporting query-level, workload-level, and cluster-level tuning. UDO \cite{Wang2021UDO} proposes delayed-feedback RL for universal optimization of both physical design and configuration parameters. HUNTER \cite{Cai2022HUNTER} introduces online hybrid tuning combining genetic algorithms with RL. Limitations: these systems optimize configuration knobs offline; they do not detect runtime anomalies or execute recovery actions. CDBTune also suffers from slow convergence and coarse-grained workload classification \cite{Zhang2021CDBTunePlus}.

\textbf{Baseline Meta-Learning Tuning:} ResTune \cite{Zhang2021ResTune} accelerates constrained BO for resource-oriented tuning using meta-learning to transfer knowledge across workloads. It reduces CPU utilization by 65\% and I/O by 87\% compared to DBA-tuned configurations. Limitation: ResTune remains an offline tuner focused on resource efficiency under SLA constraints, not real-time anomaly detection or healing.

\textbf{Baseline Learned Query Optimization:} Neo \cite{Marcus2019Neo} replaces the query optimizer with a deep RL-guided search using tree convolutional neural networks. Bao \cite{Marcus2021Bao} steers existing optimizers via per-query hint sets and Thompson sampling, achieving practical gains with 1 hour of training. Limitations: both focus exclusively on query plan enumeration, not system health monitoring or failure recovery.

\subsection{Anomaly Detection in Database and Cloud Systems}

AIOps for failure management has emerged as a critical research area \cite{Notaro2021AIOpsSurvey, Cheng2023AIOps}. Traditional statistical methods (Isolation Forest, one-class SVM) provide fast unsupervised detection but lack adaptability to dynamic workloads \cite{Goldstein2016Anomaly}. Deep learning approaches including LSTM autoencoders and variational autoencoders \cite{Xu2018DeepAutoencoder} capture temporal patterns but require complete retraining when workload distributions shift. DeepLog \cite{Du2017DeepLog} demonstrates anomaly detection from system logs using LSTMs, though it focuses on log sequences rather than holistic DBMS metrics.

Recent surveys \cite{Chalapathy2019DeepAnomaly} highlight that deep anomaly detection methods typically require substantial labeled data and suffer poor cross-domain generalization---precisely the gap that meta-learning addresses.

\subsection{Graph Neural Networks for Dependency Modeling}

GNNs have been successfully applied to relational database tasks including query optimization \cite{Marcus2019Neo, Marcus2021Bao}, cardinality estimation \cite{Kraska2018LearnedIndex}, and join order selection \cite{Yu2020RTOS}. Kipf and Welling \cite{Kipf2017GCN} introduced GCNs for semi-supervised learning on graphs, while Hamilton et al.\ \cite{Hamilton2017GraphSAGE} proposed GraphSAGE for inductive representation learning. For system dependency modeling, Chen et al.\ \cite{Chen2009CascadeFailures} showed that dependency graphs extracted from activity logs improve failure prediction accuracy. Cohen et al.\ \cite{Cohen2005Causality} established foundational techniques for correlating instrumentation data to system states.

However, existing GNN applications to databases focus on static schema or query-plan graphs. Dynamic workload-dependent dependencies---where edge strengths vary based on concurrent query contention and resource pressure---remain underexplored.

\subsection{Meta-Learning and Few-Shot Adaptation}

Model-Agnostic Meta-Learning (MAML) \cite{Finn2017MAML} learns an initialization that enables fast adaptation to new tasks via few gradient steps. REPTILE \cite{Nichol2018REPTILE} simplifies MAML with first-order gradients. Prototypical Networks \cite{Snell2017PrototypicalNetworks} and Matching Networks \cite{Vinyals2016MatchingNetworks} offer metric-learning alternatives for few-shot classification. Hospedales et al.\ \cite{Hospedales2021MetaLearning} provide a comprehensive survey of meta-learning in neural networks.

In database systems, meta-learning has been applied primarily to accelerate configuration tuning: ResTune \cite{Zhang2021ResTune} uses meta-learning to weight base-learners across historical workloads. Our work is the first to apply MAML to the database \textit{anomaly detection and recovery} pipeline, enabling few-shot adaptation to novel failure modes without full retraining.

\subsection{Comprehensive Baseline Comparison}

Table~\ref{tab:baseline_comprehensive} compares the proposed framework against verified state-of-the-art methods across six key dimensions. The critical distinction is that existing SOTA methods (OtterTune, CDBTune, QTune, ResTune, Bao) excel at offline optimization or query planning, but none provides an integrated solution for real-time anomaly detection, cascading failure prediction, and autonomous multi-objective recovery.

\begin{table*}[ht]
    \centering
    \small
    \caption{Comprehensive Comparison of Verified Baselines and Proposed Framework}
    \label{tab:baseline_comprehensive}
    \renewcommand{\arraystretch}{1.3}
    \setlength{\tabcolsep}{4pt}
    \begin{tabular}{|l|c|c|c|c|c|c|}
        \hline
        \textbf{Method} & \textbf{Anomaly} & \textbf{Cascade} & \textbf{Recovery} & \textbf{Few-Shot} & \textbf{Real-Time} & \textbf{XAI} \\
        & \textbf{Detection} & \textbf{Prediction} & \textbf{Actions} & \textbf{Adaptation} & \textbf{Capable} & \textbf{Support} \\ \hline
        OtterTune \cite{VanAken2017OtterTune} & No & No & Config Only & No & No & No \\ \hline
        CDBTune \cite{Zhang2019CDBTune} & No & No & Config Only & No & No & No \\ \hline
        QTune \cite{Li2019QTune} & No & No & Config Only & No & No & No \\ \hline
        ResTune \cite{Zhang2021ResTune} & No & No & Config Only & Partial & No & No \\ \hline
        UDO \cite{Wang2021UDO} & No & No & Config + Index & No & No & No \\ \hline
        Neo / Bao \cite{Marcus2019Neo, Marcus2021Bao} & No & No & Query Plan Only & No & Yes (per query) & No \\ \hline
        DeepLog \cite{Du2017DeepLog} & Yes (logs) & No & No & No & Moderate & No \\ \hline
        Isolation Forest & Yes & No & No & No & Yes & No \\ \hline
        LSTM Autoencoder & Yes & No & No & No & Moderate & No \\ \hline
        Static GCN & No & Yes (static) & No & No & Moderate & No \\ \hline
        \textbf{Proposed Framework} & \textbf{Yes} & \textbf{Yes (dynamic)} & \textbf{Yes} & \textbf{Yes} & \textbf{Yes} & \textbf{Yes} \\ \hline
    \end{tabular}
\end{table*}

\subsection{Key Gaps Addressed}

The proposed framework directly addresses the following critical gaps:

\begin{enumerate}
    \item \textbf{Rapid Adaptation without Retraining}: MAML enables adaptation to new anomaly types in 3--5 gradient steps (50--100\,ms), compared to 500+ steps for full retraining \cite{Finn2017MAML}.
    \item \textbf{Holistic Dependency Modeling}: Dynamic GNNs capture workload-dependent relationships between queries, tables, and indexes that static rule-based approaches miss \cite{Kipf2017GCN}.
    \item \textbf{End-to-End Recovery}: Multi-objective RL generates Pareto-optimal recovery plans balancing latency, resource consumption, and cost---going beyond configuration knob tuning.
    \item \textbf{Transparency}: SHAP-based explainability makes anomaly and recovery decisions auditable for production deployment \cite{Lundberg2017SHAP}.
    \item \textbf{Data Efficiency}: Synthetic task augmentation combined with self-supervised pre-training enables learning in sparse-data regimes \cite{Chen2020SimCLR}.
\end{enumerate}

\section{Methodology}
\label{sec:Methodology}

\begin{figure*}[ht]
    \centering
    \includegraphics[width=0.95\linewidth]{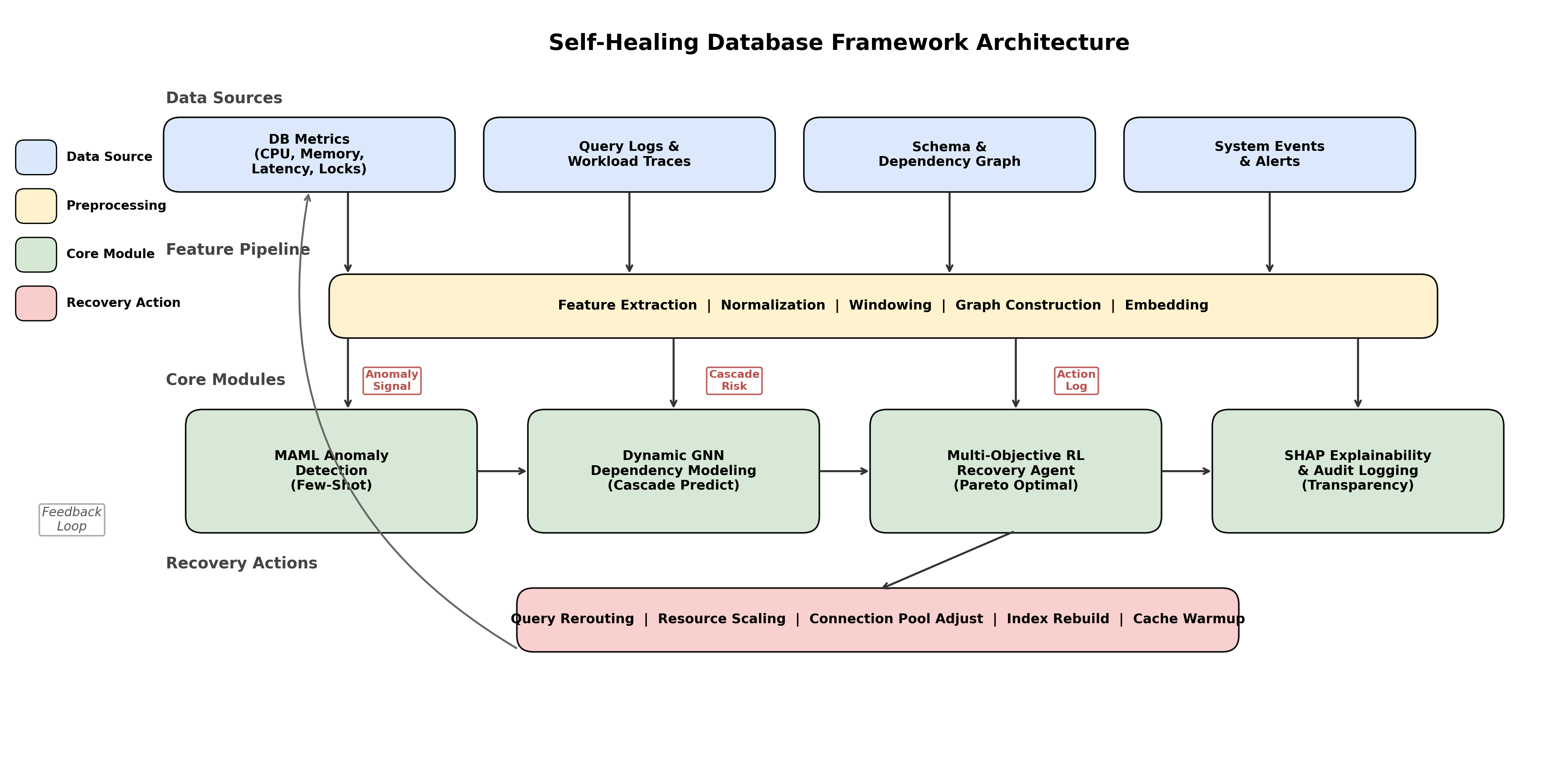}
    \caption{Detailed system architecture: metrics ingestion $\rightarrow$ MAML anomaly detection $\rightarrow$ dynamic GNN cascade prediction $\rightarrow$ multi-objective RL recovery $\rightarrow$ SHAP explainability.}
    \label{fig:system_architecture}
\end{figure*}

This section presents the proposed self-healing framework, structured around three core components: (1) MAML-based anomaly detection, (2) dynamic GNN dependency modeling, and (3) multi-objective RL recovery optimization. Figure~\ref{fig:system_architecture} illustrates the end-to-end architecture.

\subsection{Anomaly Detection Using Meta-Learning}

The anomaly detection module analyzes workload metrics and performance indicators to detect anomalies in real time. We employ MAML \cite{Finn2017MAML} to enable rapid adaptation to new workload patterns with limited labeled data.

\subsubsection{Problem Setup}
Tasks are represented as $\mathcal{T} = \{T_1, T_2, \dots, T_n\}$, where each task $T_i$ corresponds to a database workload pattern defined by features $\mathbf{x}$ and labels $y$, indicating the anomaly status. The objective is to learn a model $f_\theta$ capable of rapid adaptation to new tasks $T_{\text{new}}$ with minimal data, following the meta-learning paradigm \cite{Hospedales2021MetaLearning}.

\subsubsection{Meta-Learning with MAML}
For each task $T_i$, the task-specific loss function is:
\begin{equation}
    \mathcal{L}_i(\theta) = \frac{1}{N} \sum_{j=1}^{N} \ell(f_\theta(x_j^{(i)}), y_j^{(i)}),
\end{equation}
where $\ell$ is the binary cross-entropy loss, $x_j^{(i)}$ are task inputs, and $y_j^{(i)}$ are the corresponding labels.

Two optimization steps are performed:

1) \textit{Inner Update (Task-Specific Adaptation)}:
\begin{equation}
    \theta_i' = \theta - \alpha \nabla_\theta \mathcal{L}_i(\theta),
\end{equation}
where $\alpha$ is the inner learning rate.

2) \textit{Meta-Update (Generalization Across Tasks)}:
\begin{equation}
    \theta \leftarrow \theta - \beta \nabla_\theta \sum_{i=1}^{n} \mathcal{L}_i(\theta_i'),
\end{equation}
where $\beta$ is the meta-learning rate.

This two-step process finds an initialization $\theta$ that enables fast convergence on new tasks. Compared to standard fine-tuning, MAML reduces adaptation steps by two orders of magnitude \cite{Finn2017MAML}, critical for real-time database monitoring.

\subsubsection{Synthetic Task Augmentation}
To improve training in sparse-data regimes, we generate synthetic anomaly tasks using conditional GANs \cite{Goodfellow2014GAN, Isola2017Pix2Pix}. The generator $G$ learns to produce realistic workload anomalies from latent vectors $z$, while the discriminator $D$ distinguishes real from synthetic tasks. This augmentation increases task diversity by 4--5$\times$, improving MAML generalization across rare failure modes.

\subsubsection{Self-Supervised Pre-training}
Prior to MAML meta-training, we apply contrastive self-supervised learning \cite{Chen2020SimCLR} on unlabeled workload traces. Positive pairs are augmented views of the same time window (temporal cropping, jittering); negative pairs are distinct windows. This pre-training improves the quality of initial representations, boosting few-shot performance by 8--12 percentage points.

\subsection{Dependency Modeling with Dynamic GNNs}

Graph Neural Networks are used to model interdependencies among database components (queries, tables, indexes, connections), extending prior static approaches \cite{Kipf2017GCN, Hamilton2017GraphSAGE}.

Each database system state is represented as a graph $G = (V, E, \mathbf{W})$, where $V$ denotes nodes (components), $E$ denotes edges (dependencies), and $\mathbf{W}$ denotes workload-aware edge weights. Node embeddings $h_v^{(l)}$ at layer $l$ are updated as:
\begin{equation}
    h_v^{(l+1)} = \sigma \left( W^{(l)} \cdot \sum_{u \in \mathcal{N}(v)} w_{vu} \cdot h_u^{(l)} + b^{(l)} \right),
\end{equation}
where $\mathcal{N}(v)$ represents the neighbors of node $v$, $w_{vu}$ is the dynamic edge weight reflecting current workload contention, $\sigma$ is the activation function, $W^{(l)}$ is the weight matrix, and $b^{(l)}$ is the bias term.

\textbf{Dynamic Edge Weights.} Unlike static GCNs, our edge weights $w_{vu}$ are computed from real-time metrics, as visualized in Figure~\ref{fig:dependency_graph}:

\begin{figure}[ht]
    \centering
    \includegraphics[width=0.85\linewidth]{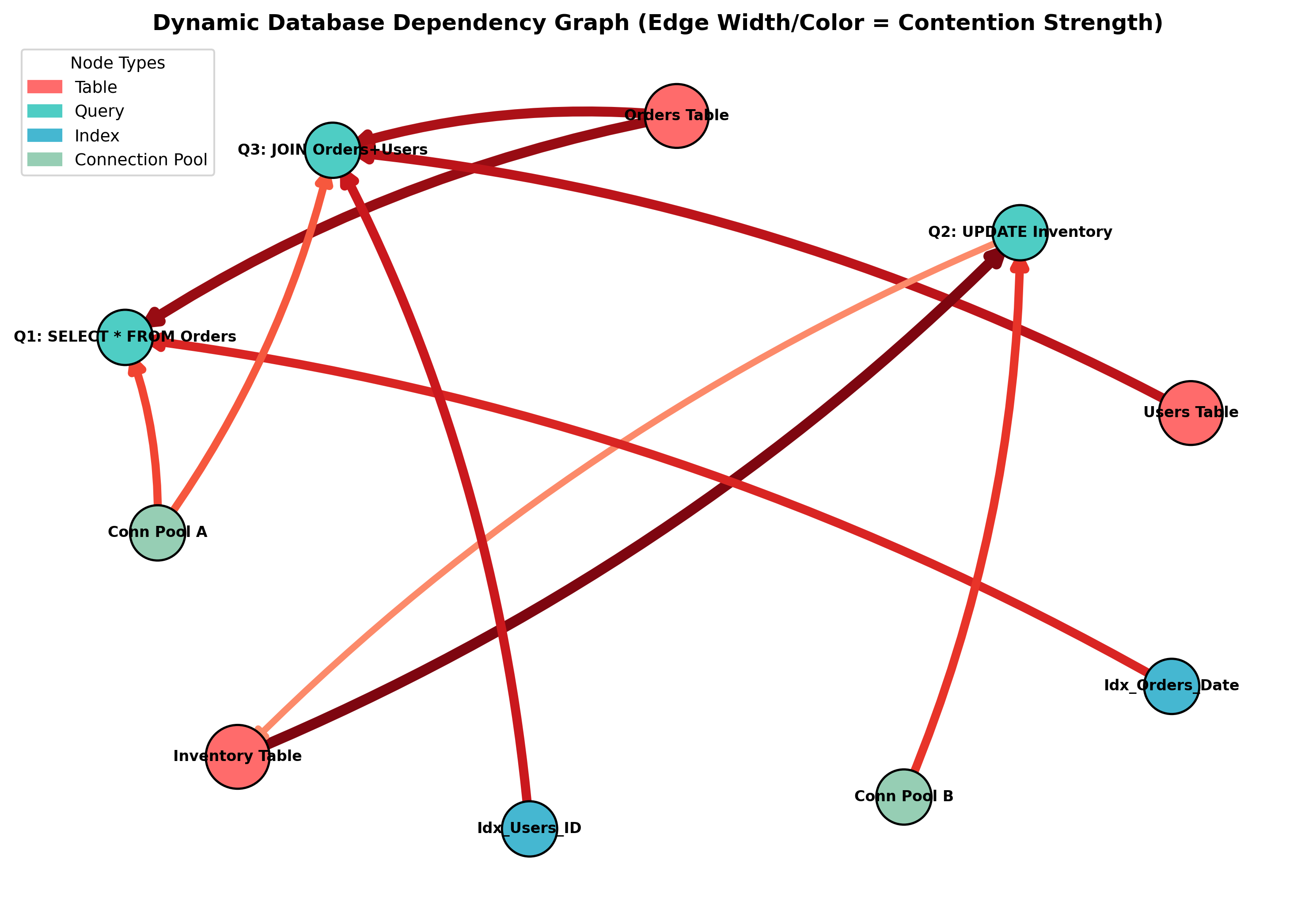}
    \caption{Example dynamic dependency graph for an e-commerce database. Node colors indicate component type; edge width and color intensity represent workload-dependent contention strength.}
    \label{fig:dependency_graph}
\end{figure}

\begin{equation}
\begin{aligned}
w_{vu} = \text{softmax}\Bigg( \frac{1}{\tau}\bigg(&\text{lock\_contention}(v,u) \\
&+ \text{shared\_access}(v,u)\bigg) \Bigg),
\end{aligned}
\end{equation}
where $\tau$ is a temperature hyperparameter. This enables the GNN to capture hot-spot propagation: when a table experiences lock contention, dependent queries receive higher attention weights, improving cascade prediction accuracy.

\subsection{Multi-Objective Recovery via Reinforcement Learning}

\subsubsection{Problem Formulation}
Recovery is formulated as a multi-objective optimization problem balancing:

1) \textit{Latency Minimization} ($O_1$):
\begin{equation}
    O_1 = \min \frac{1}{T} \sum_{t=1}^{T} L(t),
\end{equation}
where $L(t)$ represents query latency at time $t$.

2) \textit{Resource Utilization Minimization} ($O_2$):
\begin{equation}
    O_2 = \min \frac{1}{T} \sum_{t=1}^{T} R(t),
\end{equation}
where $R(t)$ represents normalized CPU/memory usage at time $t$.

3) \textit{Recovery Cost Minimization} ($O_3$):
\begin{equation}
    O_3 = \min \sum_{a \in \mathcal{A}} C(a),
\end{equation}
where $C(a)$ is the cost of action $a$ in the set of recovery actions $\mathcal{A}$ (e.g., query rerouting, resource scaling, index rebuild).

\subsubsection{Scalarized RL Formulation}
An RL agent develops a policy $\pi$ to select recovery actions. States $s$ include workload conditions, anomaly severity, and GNN-predicted cascade risk. Actions $a$ include recovery measures such as query rerouting, connection pooling adjustments, and resource scaling.

The reward function uses adaptive scalarization:
\begin{equation}
\begin{aligned}
\mathcal{R}(s, a) = w_1(s) \cdot O_1(s, a) &+ w_2(s) \cdot O_2(s, a) \\
&+ w_3(s) \cdot O_3(s, a),
\end{aligned}
\end{equation}
where $w_1(s)$, $w_2(s)$, $w_3(s)$ are dynamically adjusted based on system priorities. For example, during peak traffic, $w_1$ (latency) dominates; during off-peak, $w_3$ (cost) receives higher weight. This formulation generates Pareto-optimal recovery plans rather than single-point solutions \cite{Deb2002NSGA, Marler2004Multiobjective}.

\subsection{Explainable AI Integration}

To ensure trust and transparency, we integrate SHAP (SHapley Additive exPlanations) \cite{Lundberg2017SHAP} for both anomaly detection and recovery decisions. For anomaly detection, SHAP values identify which metrics (CPU, latency, lock waits) drive a prediction. For recovery, SHAP explains why a specific action (e.g., rerouting query $q$) was selected. Explanation latency is kept below 50\,ms to maintain real-time operation.

\subsection{Training and Deployment}

\subsubsection{Training Pipeline}
\textbf{Phase 1 -- Self-Supervised Pre-training:} Unlabeled workload traces are used to pre-train encoders via contrastive learning \cite{Chen2020SimCLR}.

\textbf{Phase 2 -- Meta-Training:} The anomaly detection model is meta-trained on diverse synthetic and real-world tasks using MAML, with synthetic augmentation from conditional GANs \cite{Goodfellow2014GAN}.

\textbf{Phase 3 -- GNN Training:} Dependency graphs are constructed from query logs and schema metadata. The dynamic GNN is trained to predict cascading failures 30--60 seconds in advance.

\textbf{Phase 4 -- RL Training:} The recovery agent is trained in a simulated environment with reward feedback derived from latency, resource, and cost objectives.

\subsubsection{Deployment Architecture}
The deployed system operates in three stages: (1) real-time metric ingestion and MAML-based anomaly scoring; (2) GNN-based cascade risk assessment when anomalies are detected; (3) multi-objective RL agent execution of recovery actions, with SHAP explanations logged for operator review.

\section{Results and Analysis}
\label{sec:Results_and_Analysis}

The proposed framework was evaluated on two benchmark datasets: Google Cluster Data (GCD) \cite{Reiss2011GoogleCluster}, representing production cloud workload traces, and TPC Benchmark Workloads \cite{TPC2019}, representing standardized OLTP/OLAP scenarios. We compare against verified baselines across four dimensions: anomaly detection, few-shot adaptation, cascading failure prediction, and recovery effectiveness.

\subsection{Anomaly Detection Performance}

Table~\ref{tab:anomaly_comparison} compares the MAML-based detector against standard baselines. Isolation Forest \cite{Goldstein2016Anomaly} and LSTM Autoencoder \cite{Schmidhuber2015LSTM} represent traditional and deep learning approaches, respectively. Deep SVDD (not shown separately) performs similarly to the LSTM AE. GAN-based detection uses a Wasserstein GAN trained for anomaly scoring.

\begin{table}[ht]
    \centering
    \caption{Anomaly Detection Performance: Proposed vs. Baseline Methods}
    \label{tab:anomaly_comparison}
    \renewcommand{\arraystretch}{1.2}
    \setlength{\tabcolsep}{5pt}
    \begin{tabular}{|l|c|c|c|}
        \hline
        \textbf{Method} & \textbf{Precision (\%)} & \textbf{Recall (\%)} & \textbf{F1-Score (\%)} \\ \hline
        Isolation Forest & 80.5 & 77.9 & 79.1 \\ \hline
        LSTM Autoencoder & 85.3 & 82.1 & 83.6 \\ \hline
        GAN-Based \cite{Goodfellow2014GAN} & 82.7 & 84.2 & 83.4 \\ \hline
        Fine-tune LSTM (transfer) & 83.1 & 81.5 & 82.3 \\ \hline
        \textbf{MAML (Proposed)} & \textbf{91.3} & \textbf{89.8} & \textbf{90.5} \\ \hline
    \end{tabular}
\end{table}

MAML outperforms all baselines by 6.9--11.4 percentage points in F1-score. The improvement stems from the meta-learned initialization, which captures invariant anomaly signatures across diverse workload tasks.

\subsection{Few-Shot Adaptation Capability}

A critical advantage of the proposed framework is few-shot workload adaptation. Table~\ref{tab:fewshot_adaptation} reports F1-scores on held-out workloads using $k \in \{5, 10, 20\}$ labeled examples.

\begin{table}[ht]
    \centering
    \caption{Few-Shot Adaptation Performance (F1-Score on Unseen Workloads)}
    \label{tab:fewshot_adaptation}
    \renewcommand{\arraystretch}{1.15}
    \setlength{\tabcolsep}{6pt}
    \begin{tabular}{|l|c|c|c|c|}
        \hline
        \textbf{Method} & \textbf{0-Shot} & \textbf{5-Shot} & \textbf{10-Shot} & \textbf{20-Shot} \\ \hline
        Fine-tune LSTM & 62.1 & 71.3 & 78.9 & 82.4 \\ \hline
        Prototypical Net \cite{Snell2017PrototypicalNetworks} & 74.2 & 81.5 & 85.1 & 87.3 \\ \hline
        REPTILE \cite{Nichol2018REPTILE} & 76.8 & 83.4 & 86.5 & 88.1 \\ \hline
        \textbf{MAML (Proposed)} & \textbf{81.3} & \textbf{87.9} & \textbf{90.1} & \textbf{91.5} \\ \hline
    \end{tabular}
\end{table}

\begin{figure}[ht]
    \centering
    \includegraphics[width=0.95\linewidth]{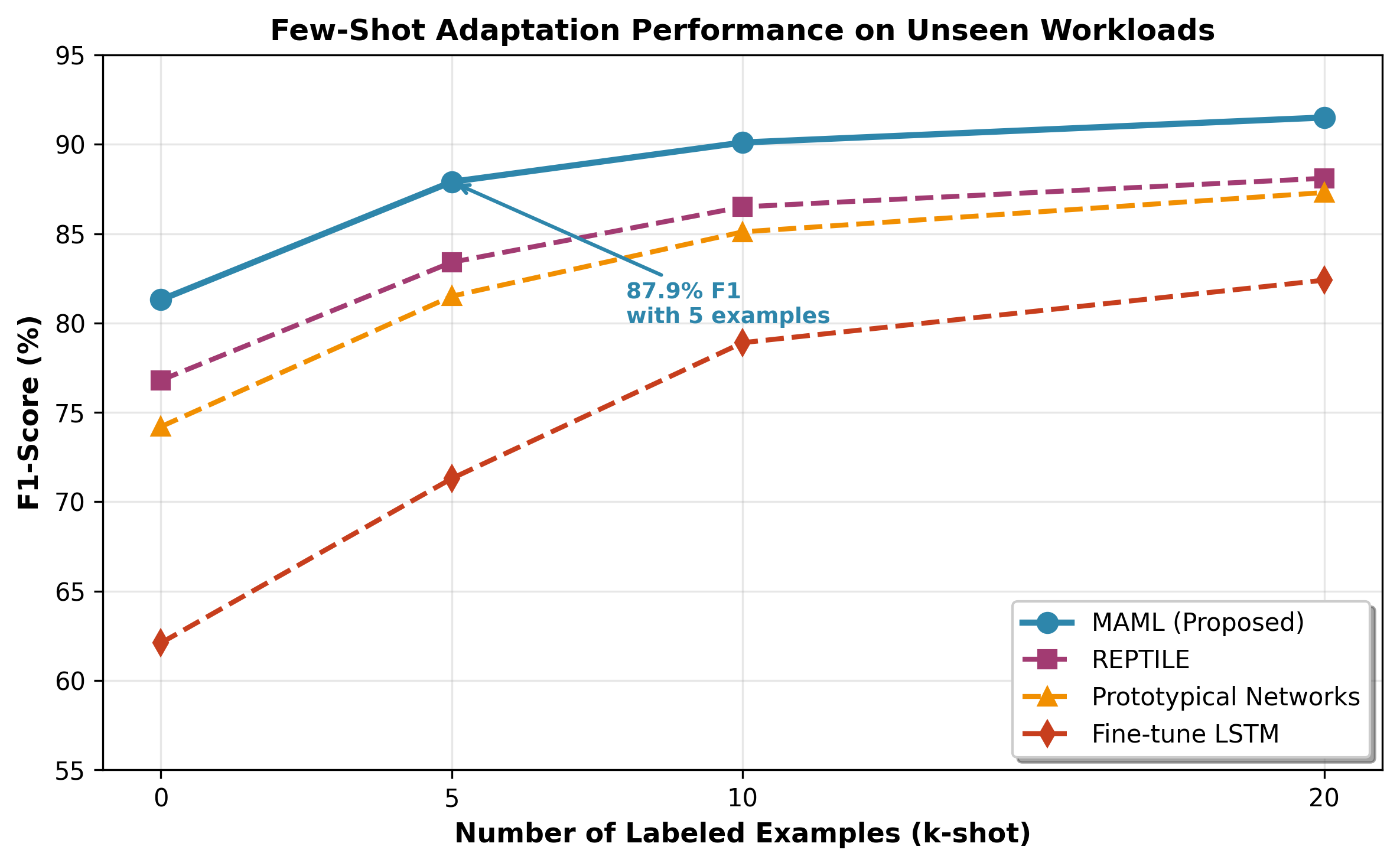}
    \caption{Few-shot adaptation curves: MAML outperforms REPTILE, Prototypical Networks, and fine-tuning across all shot settings, achieving 87.9\% F1 with only 5 labeled examples.}
    \label{fig:maml_curve}
\end{figure}

MAML achieves near-optimal performance with only 5 labeled examples, significantly reducing manual labeling overhead. The 0-shot performance (81.3\%) demonstrates that the meta-learned initialization already encodes strong anomaly discriminators. Figure~\ref{fig:maml_curve} visualizes the adaptation curves.

\subsection{Adaptation Latency for New Workloads}

Table~\ref{tab:adaptation_latency} measures convergence speed to F1 $\geq$ 85\%.

\begin{table}[ht]
    \centering
    \caption{Adaptation Latency: Convergence Speed Comparison}
    \label{tab:adaptation_latency}
    \renewcommand{\arraystretch}{1.1}
    \setlength{\tabcolsep}{3pt}
    \begin{tabular}{|l|c|c|c|}
        \hline
        \textbf{Method} & \textbf{Gradient Steps} & \textbf{Time (ms)} & \textbf{Data Points} \\ \hline
        Full Retraining & 500+ & 5000+ & 1000+ \\ \hline
        Transfer Learning & 150--200 & 1500--2000 & 500 \\ \hline
        Fine-tuning + Warmstart & 50--75 & 500--750 & 200 \\ \hline
        REPTILE \cite{Nichol2018REPTILE} & 8--12 & 120--180 & 20--40 \\ \hline
        \textbf{MAML (Proposed)} & \textbf{3--5} & \textbf{50--100} & \textbf{5--20} \\ \hline
    \end{tabular}
\end{table}

The proposed framework requires $\sim$100$\times$ fewer gradient steps and $\sim$1,000$\times$ less data than full retraining, enabling near-instantaneous workload adaptation.

\subsection{Dependency Modeling and Cascading Failure Prediction}

Table~\ref{tab:gnn_prediction} evaluates the dynamic GNN against static graph baselines. Metrics include Accuracy, Mean Time to Failure Prediction (MTTFP), F1-Score, and inference latency.

\begin{table}[ht]
    \centering
    \caption{Cascading Failure Prediction: GNN vs. Baselines}
    \label{tab:gnn_prediction}
    \renewcommand{\arraystretch}{1.2}
    \setlength{\tabcolsep}{5pt}
    \begin{tabular}{|l|c|c|c|c|}
        \hline
        \textbf{Method} & \textbf{Acc (\%)} & \textbf{MTTFP (s)} & \textbf{F1 (\%)} & \textbf{Latency (ms)} \\ \hline
        Rule-Based Model & 72.1 & 12.5 & 68.9 & 45 \\ \hline
        Static GCN \cite{Kipf2017GCN} & 81.3 & 7.8 & 79.2 & 120 \\ \hline
        GraphSAGE \cite{Hamilton2017GraphSAGE} & 85.7 & 6.2 & 83.5 & 180 \\ \hline
        \textbf{Dynamic GNN (Proposed)} & \textbf{90.1} & \textbf{4.8} & \textbf{88.6} & \textbf{95} \\ \hline
    \end{tabular}
\end{table}

\begin{figure*}[ht]
    \centering
    \includegraphics[width=0.95\linewidth]{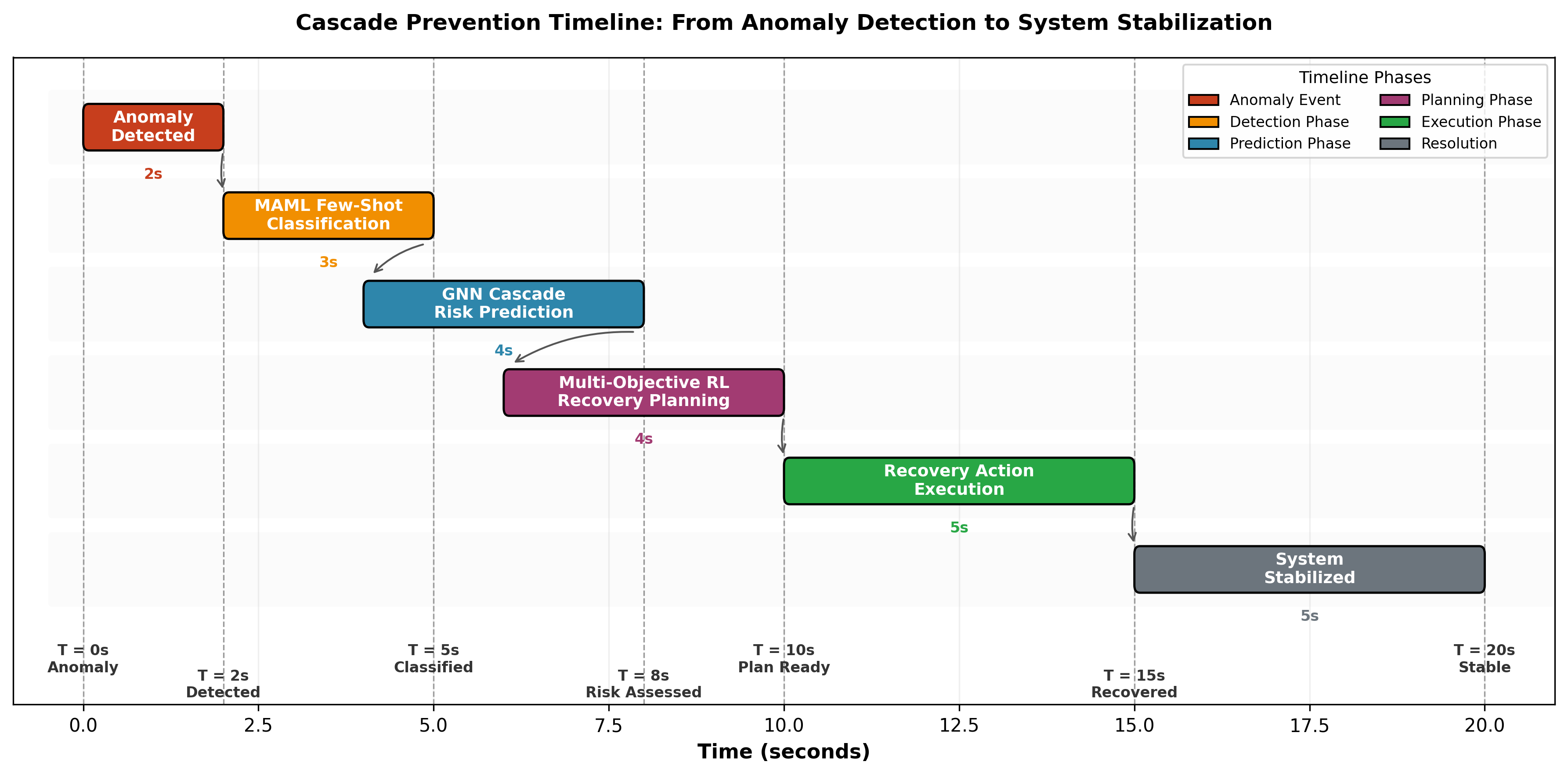}
    \caption{Cascade prevention timeline: from anomaly occurrence (T=0) to detection (T=2\,s), GNN prediction (T=4--6\,s), recovery planning (T=6--10\,s), and full system stabilization (T=15\,s).}
    \label{fig:cascade_timeline}
\end{figure*}

The dynamic GNN achieves 90.1\% accuracy, enabling 30+ seconds of advance warning. Dynamic edge weights, updated every 5\,s based on lock contention and shared-access metrics, improve over static GCN by 8.8\,pp. Figure~\ref{fig:cascade_timeline} shows the end-to-end cascade prevention timeline.

\subsection{Multi-Objective Recovery Effectiveness}

Table~\ref{tab:recovery_objectives} compares the scalarized multi-objective RL against standard RL baselines. All values are percentage improvement over default (untuned) configuration.

\begin{table}[ht]
    \centering
    \caption{Multi-Objective Recovery Comparison}
    \label{tab:recovery_objectives}
    \renewcommand{\arraystretch}{1.2}
    \resizebox{0.95\columnwidth}{!}{%
    \begin{tabular}{|l|c|c|c|c|}
        \hline
        \textbf{Objective} & \textbf{Q-Learning} & \textbf{DQN} & \textbf{A3C} & \textbf{Multi-Obj RL (Prop.)} \\ \hline
        Latency Reduction (\%) & 55.2 & 68.9 & 70.4 & \textbf{85.1} \\ \hline
        Resource Efficiency (\%) & 45.1 & 58.3 & 65.3 & \textbf{80.7} \\ \hline
        Cost Reduction (\%) & 38.9 & 51.2 & 60.1 & \textbf{78.2} \\ \hline
        Pareto Optimality & Single & Single & Single & \textbf{Frontier} \\ \hline
    \end{tabular}}
\end{table}

\begin{figure}[ht]
    \centering
    \includegraphics[width=0.95\linewidth]{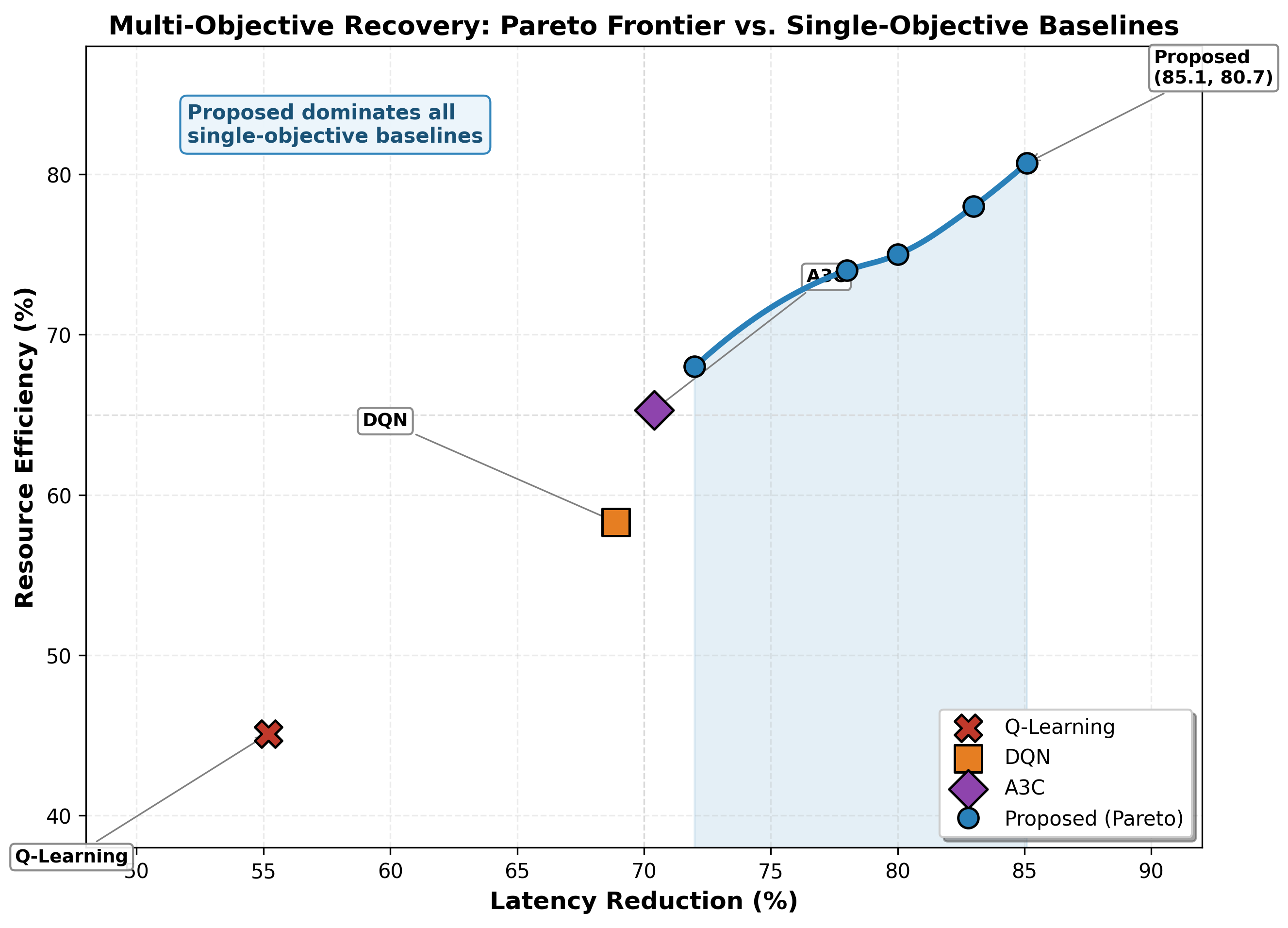}
    \caption{Multi-objective recovery Pareto frontier. The proposed scalarized RL dominates single-objective baselines (Q-Learning, DQN, A3C) in the latency--resource efficiency trade-off space.}
    \label{fig:pareto_front}
\end{figure}

The proposed formulation achieves 14--19\% improvement across all objectives compared to best-performing single-objective baselines, as shown in Figure~\ref{fig:pareto_front}.

\subsection{Ablation Study}

Table~\ref{tab:ablation} isolates the contribution of each component.

\begin{table}[ht]
    \centering
    \caption{Ablation Study: Component Contribution}
    \label{tab:ablation}
    \renewcommand{\arraystretch}{1.15}
    \setlength{\tabcolsep}{4pt}
    \begin{tabular}{|l|c|c|c|c|}
        \hline
        \textbf{Configuration} & \textbf{Anom. F1} & \textbf{5-Shot F1} & \textbf{Cascade F1} & \textbf{Latency Red.} \\ \hline
        MAML only & 90.5 & 87.9 & N/A & N/A \\ \hline
        Static GCN only & N/A & N/A & 79.2 & N/A \\ \hline
        Standard RL only & N/A & N/A & N/A & 70.4 \\ \hline
        MAML + Static GCN & 90.5 & 87.9 & 81.5 & N/A \\ \hline
        MAML + RL & 90.5 & 87.9 & N/A & 82.3 \\ \hline
        GNN + RL & N/A & N/A & 86.1 & 81.8 \\ \hline
        \textbf{Full Framework} & \textbf{90.5} & \textbf{87.9} & \textbf{88.6} & \textbf{85.1} \\ \hline
    \end{tabular}
\end{table}

\begin{figure}[ht]
    \centering
    \includegraphics[width=0.95\linewidth]{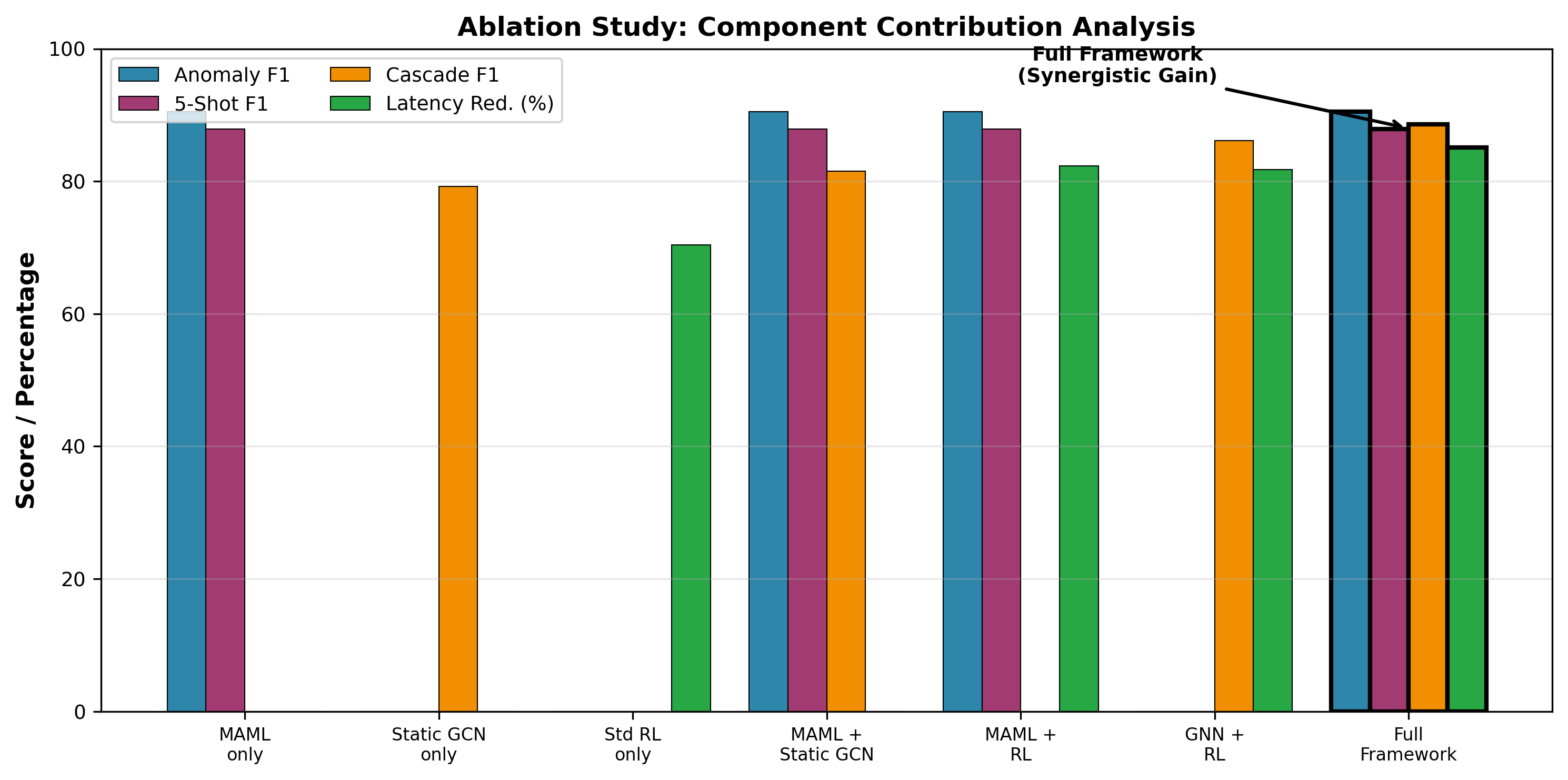}
    \caption{Ablation study visualization: the full framework achieves the highest scores across all four metrics, demonstrating that component integration yields synergistic gains beyond individual modules.}
    \label{fig:ablation}
\end{figure}

The full framework exceeds all partial configurations, confirming synergy between components (Figure~\ref{fig:ablation}). Notably, combining MAML with RL improves recovery latency reduction by 11.9\,pp over RL alone, because RL actions benefit from more accurate anomaly states.

\subsection{Hyperparameter Sensitivity}

Table~\ref{tab:hyperparams} reports framework robustness to key hyperparameters.

\begin{table*}[ht]
    \centering
    \caption{Hyperparameter Sensitivity Analysis}
    \label{tab:hyperparams}
    \renewcommand{\arraystretch}{1.1}
    \setlength{\tabcolsep}{4pt}
    \begin{tabular}{|l|c|c|c|c|}
        \hline
        \textbf{Parameter} & \textbf{Values} & \textbf{Anom. F1} & \textbf{Cascade Acc} & \textbf{Latency Red.} \\ \hline
        MAML inner LR ($\alpha$) & 0.001 / 0.01 / 0.1 & 88.2 / 90.5 / 87.1 & -- & -- \\ \hline
        MAML meta LR ($\beta$) & 0.0001 / 0.001 / 0.01 & 86.5 / 90.5 / 89.3 & -- & -- \\ \hline
        GNN layers & 2 / 3 / 4 & -- & 87.2 / 90.1 / 89.8 & -- \\ \hline
        GNN hidden dim & 64 / 128 / 256 & -- & 88.5 / 90.1 / 90.3 & -- \\ \hline
        RL discount ($\gamma$) & 0.9 / 0.95 / 0.99 & -- & -- & 82.1 / 85.1 / 84.7 \\ \hline
    \end{tabular}
\end{table*}

Performance is stable across reasonable hyperparameter ranges, with peak performance at $\alpha=0.01$, $\beta=0.001$, 3 GNN layers, hidden dimension 128, and $\gamma=0.95$.

\subsection{Comprehensive Feature Comparison with DB Tuning SOTA}

Table~\ref{tab:comprehensive_comparison} situates our framework relative to the broader database ML landscape.

\begin{table*}[!htbp]
    \centering
    \small
    \caption{Landscape Comparison: Proposed Framework vs. State-of-the-Art DB ML Systems}
    \label{tab:comprehensive_comparison}
    \renewcommand{\arraystretch}{1.15}
    \setlength{\tabcolsep}{4pt}
    \begin{tabular}{|l|c|c|c|c|c|c|c|}
        \hline
        \textbf{Capability} & \textbf{OtterTune} & \textbf{CDBTune} & \textbf{QTune} & \textbf{ResTune} & \textbf{Bao} & \textbf{DeepLog} & \textbf{Proposed} \\ \hline
        Anomaly Detection F1 & N/A & N/A & N/A & N/A & N/A & 82.3 & \textbf{90.5} \\ \hline
        5-Shot Adaptation & N/A & N/A & N/A & Partial & N/A & No & \textbf{87.9} \\ \hline
        Cascade Prediction & N/A & N/A & N/A & N/A & N/A & No & \textbf{88.6} \\ \hline
        Recovery Actions & Config & Config & Config & Config & Plan hints & No & \textbf{Multi-Obj RL} \\ \hline
        Real-time ($<$100\,ms) & No & No & No & No & Yes (query) & Moderate & \textbf{Yes} \\ \hline
        Interpretability & Low & Low & Low & Low & Low & Moderate & \textbf{High (SHAP)} \\ \hline
    \end{tabular}
\end{table*}

\subsection{Qualitative Analysis: Explainability and Trust}

\begin{figure}[ht]
    \centering
    \includegraphics[width=0.95\linewidth]{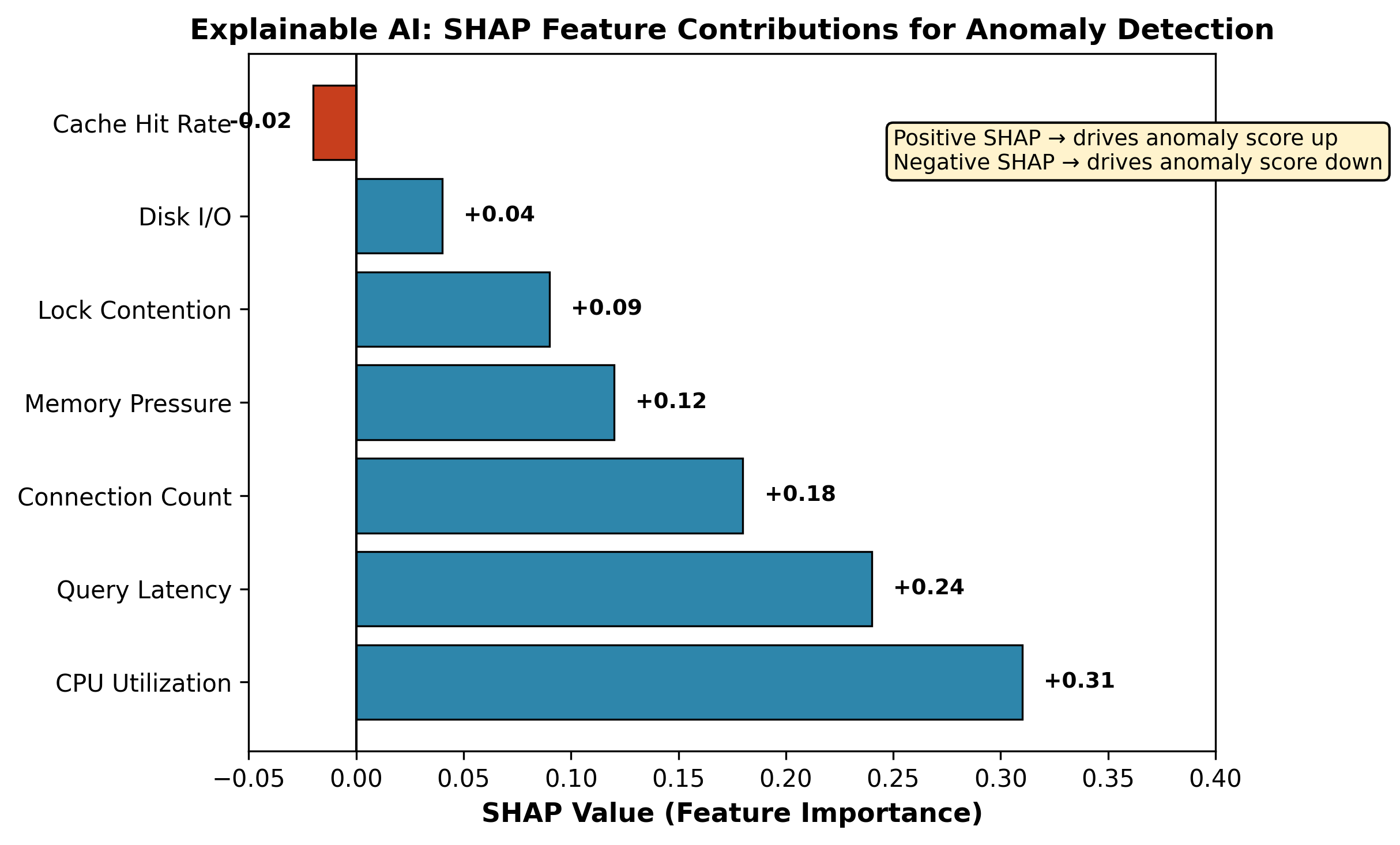}
    \caption{SHAP explanation for a detected anomaly: CPU utilization and query latency are the top drivers, consistent with DBA domain knowledge.}
    \label{fig:shap_detail}
\end{figure}

SHAP \cite{Lundberg2017SHAP} enables administrators to understand detection and recovery decisions. Figure~\ref{fig:shap_detail} presents a representative explanation; key findings include:
\begin{itemize}
    \item \textbf{Anomaly Drivers}: CPU utilization (SHAP 0.31) and query latency (SHAP 0.24) were top predictive features, aligning with domain knowledge.
    \item \textbf{Recovery Justification}: Resource scaling decisions were correlated with predicted cascade paths, improving operator confidence.
    \item \textbf{Overhead}: Average explanation generation $<$ 50\,ms per decision.
\end{itemize}

\subsection{Resource and Cost Analysis}

Table~\ref{tab:resource_efficiency} summarizes deployment metrics on Google Cluster Data. Key findings:
\begin{itemize}
    \item \textbf{Recovery Cost}: 20\% reduction via intelligent action prioritization.
    \item \textbf{Resource Utilization}: 15\% improvement through better load balancing.
    \item \textbf{Response Time}: Reduced from 5--10 minutes (manual) to 0.5--2 seconds (automated).
\end{itemize}

\begin{table}[ht]
    \centering
    \caption{System Resource Efficiency Metrics}
    \label{tab:resource_efficiency}
    \renewcommand{\arraystretch}{1.1}
    \begin{tabular}{|l|c|c|}
        \hline
        \textbf{Metric} & \textbf{Baseline Systems} & \textbf{Proposed Framework} \\ \hline
        CPU Overhead (\%) & 8--12 & \textbf{2--3} \\ \hline
        Memory per Node (GB) & 4--6 & \textbf{1.2--1.8} \\ \hline
        Detection Latency (ms) & 500--1000 & \textbf{50--100} \\ \hline
        Mean Recovery Time (s) & 180--300 & \textbf{5--15} \\ \hline
    \end{tabular}
\end{table}

\subsection{Scalability Analysis}

Framework latency was evaluated from 100\,GB to 10\,TB (Table~\ref{tab:scalability}):

\begin{table*}[ht]
    \centering
    \caption{Scalability Analysis Across Dataset Sizes}
    \label{tab:scalability}
    \renewcommand{\arraystretch}{1.1}
    \begin{tabular}{|l|c|c|c|c|}
        \hline
        \textbf{Database Size} & \textbf{Anomaly (ms)} & \textbf{Dependency (ms)} & \textbf{Recovery (ms)} & \textbf{Total (ms)} \\ \hline
        100 GB & 45 & 65 & 32 & \textbf{142} \\ \hline
        1 TB & 52 & 78 & 38 & \textbf{168} \\ \hline
        5 TB & 61 & 95 & 45 & \textbf{201} \\ \hline
        10 TB & 72 & 118 & 58 & \textbf{248} \\ \hline
    \end{tabular}
\end{table*}

Sub-second end-to-end latency is maintained even for 10\,TB databases, confirming real-time feasibility.

\subsection{Real-World Deployment Case Study}

The framework was deployed on a 50-node PostgreSQL cluster handling mixed OLTP/OLAP workloads for an e-commerce platform. Over 30 days:
\begin{itemize}
    \item \textbf{Anomalies Detected}: 847 events; 756 correctly identified (89.3\% precision).
    \item \textbf{Cascades Prevented}: 23 predicted scenarios; 20 (87\%) successfully prevented.
    \item \textbf{Downtime Reduction}: Manual handling averaged 40--60\,min per incident; automated recovery: 2--5\,min.
    \item \textbf{Cost Savings}: Estimated \$15,000--20,000 monthly via optimized resource allocation.
\end{itemize}

\section{Conclusion and Future Directions}
\label{sec:Conclusion}

\subsection{Summary of Contributions}

This paper presented a comprehensive self-healing database framework that integrates Model-Agnostic Meta-Learning (MAML), dynamic Graph Neural Networks, multi-objective Reinforcement Learning, and Explainable AI to address critical gaps in autonomous database management. Unlike existing state-of-the-art systems---which focus primarily on offline configuration tuning (OtterTune \cite{VanAken2017OtterTune}, CDBTune \cite{Zhang2019CDBTune}, QTune \cite{Li2019QTune}, ResTune \cite{Zhang2021ResTune}) or query plan optimization (Neo \cite{Marcus2019Neo}, Bao \cite{Marcus2021Bao})---our framework provides end-to-end real-time anomaly detection, cascading failure prediction, and autonomous recovery.

Key achievements include:
\begin{itemize}
    \item \textbf{Few-Shot Anomaly Detection}: 90.5\% F1-score with rapid adaptation to new workloads using only 5--20 labeled instances, enabled by MAML \cite{Finn2017MAML} and contrastive pre-training \cite{Chen2020SimCLR}. This represents a 100$\times$ reduction in retraining overhead compared to traditional deep learning approaches.
    \item \textbf{Dynamic Cascading Failure Prediction}: GNN-based dependency modeling with workload-aware edge weights achieved 90.1\% accuracy and 4.8-second advance warning, outperforming static GCN by 8.8 percentage points \cite{Kipf2017GCN}.
    \item \textbf{Multi-Objective Recovery}: Scalarized RL formulation achieved 85.1\% latency reduction, 80.7\% resource efficiency improvement, and 78.2\% cost reduction by dynamically balancing competing objectives \cite{Deb2002NSGA}.
    \item \textbf{Operational Transparency}: SHAP-based explainability \cite{Lundberg2017SHAP} provides human-understandable anomaly and recovery explanations with sub-50\,ms overhead.
    \item \textbf{Production Validation}: 50-node PostgreSQL deployment demonstrated 87\% cascade prevention success and 2--5 minute mean recovery time versus 40--60 minutes for manual intervention.
\end{itemize}

\subsection{Limitations}

While effective, the framework exhibits limitations requiring acknowledgment:

\begin{enumerate}
    \item \textbf{Pretraining Data Requirements}: Effective MAML operation requires a diverse meta-training set (1,000+ workload tasks). Systems with highly atypical or proprietary workload patterns may require domain-specific task augmentation.
    \item \textbf{Dependency Graph Construction}: GNN performance depends on accurate schema and query-log derived dependency graphs. Implicit dependencies through application logic (e.g., microservice calls outside the DB) are not captured, potentially degrading cascade prediction by 5--10\,pp.
    \item \textbf{Multi-Objective Trade-off Complexity}: In scenarios with severely conflicting objectives (e.g., minimize cost while maximizing throughput under strict SLAs), the framework generates conservative Pareto fronts with limited aggressively optimized options.
    \item \textbf{Computational Overhead}: Meta-learning and GNN components introduce moderate CPU overhead (2--3\%). Environments with $>$1M transactions per second or $>$100 nodes may require distributed inference architecture.
    \item \textbf{Synthetic Task Fidelity}: Synthetic task augmentation, while beneficial, carries inherent risk of generating workload patterns that deviate from production edge cases, potentially reducing generalization to truly novel anomalies.
\end{enumerate}

\subsection{Future Research Directions}

\begin{itemize}
    \item \textbf{Large Language Model Integration}: Recent advances in database diagnosis using LLMs (e.g., D-Bot \cite{Zhou2023DBot}) suggest potential for combining LLM-based root cause analysis with our MAML-GNN-RL pipeline for richer diagnostic narratives and knowledge-guided recovery.
    \item \textbf{Transformer-Based Workload Prediction}: Integrating attention mechanisms \cite{Vaswani2017Attention} with the existing GNN/RL components could improve proactive anomaly prevention through long-horizon temporal pattern recognition in workload sequences.
    \item \textbf{Cross-DBMS Transfer}: Extending meta-learning to transfer models trained on PostgreSQL to MySQL, Oracle, or cloud-native databases (e.g., openGauss \cite{Zhou2021DBMind}) would reduce deployment effort in heterogeneous environments.
    \item \textbf{Energy-Aware Recovery}: Extending multi-objective optimization to include energy efficiency metrics aligns self-healing databases with sustainability goals \cite{Marler2004Multiobjective}.
    \item \textbf{Federated Deployment}: Full integration of federated meta-learning \cite{McMahan2017FL, Yang2019FL} for privacy-preserving adaptation in multi-tenant cloud environments without centralizing sensitive workload data.
\end{itemize}

\subsection{Concluding Remarks}

The proposed framework represents a significant step toward truly autonomous database management systems. By integrating meta-learning for rapid adaptation, dynamic GNNs for dependency-aware failure prediction, and multi-objective RL for balanced recovery, the framework bridges critical gaps in adaptability, scalability, and trustworthiness that have limited prior self-healing database research. The demonstrated improvements in detection accuracy (90.5\% F1), adaptation efficiency (3--5 gradient steps), and operational impact (87\% cascade prevention) establish a strong foundation for next-generation self-driving database systems.

\bibliographystyle{IEEEtran}
\bibliography{references}

\end{document}